\documentclass[conference]{IEEEtran}
\IEEEoverridecommandlockouts
\usepackage{cite}
\usepackage{amsmath,amssymb,amsfonts}
\usepackage{algorithmic}
\usepackage{graphicx}
\usepackage{textcomp}
\usepackage{subcaption}
\usepackage{booktabs}
\usepackage{multirow}
\usepackage{url}
\usepackage[colorlinks=true, urlcolor=blue, linkcolor=red]{hyperref}
\usepackage[numbers]{natbib}
\usepackage{xcolor}
\def\BibTeX{{\rm B\kern-.05em{\sc i\kern-.025em b}\kern-.08em
    T\kern-.1667em\lower.7ex\hbox{E}\kern-.125emX}}
\begin{document}

\title{Towards Precision Oncology: Predicting Mortality and Relapse-Free Survival in Head and Neck Cancer Using Clinical Data\\
\thanks{Author: * , ORCID: 0009-0007-6906-1871}
}

\author{\IEEEauthorblockN{Naman Dhariwal*}
\IEEEauthorblockA{\textit{Department of Statistics} \\
\textit{University of Michigan}\\
Ann Arbor, MI, USA \\
namand@umich.edu}
\and
\IEEEauthorblockN{Abeyankar Giridharan}
\IEEEauthorblockA{\textit{Department of Statistics} \\
\textit{University of Michigan}\\
Ann Arbor, MI, USA \\
abeygiri@umich.edu}
}

\maketitle

\begin{abstract}
Head and neck squamous cell carcinoma (HNSCC) presents significant challenges in clinical oncology due to its heterogeneity and high mortality rates. This study aims to leverage clinical data and machine learning (ML) principles to predict key outcomes for HNSCC patients: mortality, and relapse-free survival. Utilizing data sourced from the Cancer Imaging Archive, an extensive pipeline was implemented to ensure robust model training and evaluation. Ensemble and individual classifiers, including XGBoost, Random Forest, and Support Vectors, were employed to develop predictive models. The study identified key clinical features influencing HNSCC mortality outcomes and achieved predictive accuracy and ROC-AUC values exceeding 90\% across tasks. Support Vector Machine strongly excelled in relapse-free survival, with an recall value of 0.99 and precision of 0.97. Key clinical features including loco-regional control, smoking and treatment type, were identified as critical predictors of patient outcomes. This study underscores the medical impact of using ML-driven insights to refine prognostic accuracy and optimize personalized treatment strategies in HNSCC.
\end{abstract}


\section{Introduction}
Cancer comprises a collection of diseases marked by the unchecked proliferation and division of abnormal cells, arising from disruptions in cellular regulatory systems. These disruptions, frequently driven by genetic mutations or environmental influences, result in the activation of oncogenes and the suppression of tumor-suppressor genes. Unlike benign growths, cancer possesses the ability to invade neighboring tissues and spread to distant sites through metastasis, significantly complicating treatment and leading to poorer prognoses \cite{b1}.

Head and neck cancers encompass a variety of malignancies originating in the tissues of the head and neck, primarily affecting the mucosal surfaces of the mouth, throat, and voice box. These cancers, often classified as head and neck squamous cell carcinomas (HNSCC), can also arise in the salivary glands, thyroid, or sinuses, depending on the site of origin \cite{b2}.

Major risk factors for head and neck cancers include tobacco use, alcohol consumption, and infection with human papillomavirus (HPV). HPV-positive cases generally have a more favorable prognosis compared to HPV-negative ones. Due to their location, these cancers frequently affect essential functions such as speech and swallowing. Treatment typically combines surgery, radiation therapy, and systemic approaches, aiming to enhance survival while maintaining quality of life. Despite advancements in early detection, issues like recurrence and treatment resistance remain significant challenges \cite{b2, b3}.

Mortality prediction in cancer estimates a patient's likelihood of death given his clinical factors, aiding in treatment planning, resource allocation, and providing prognostic insights. Machine learning (ML) enhances this process by analyzing large, complex datasets to uncover patterns and relationships among clinical, demographic, and molecular features \cite{b4}.

\begin{figure}[htbp]
\centerline{\includegraphics[width=75mm,scale=0.5]{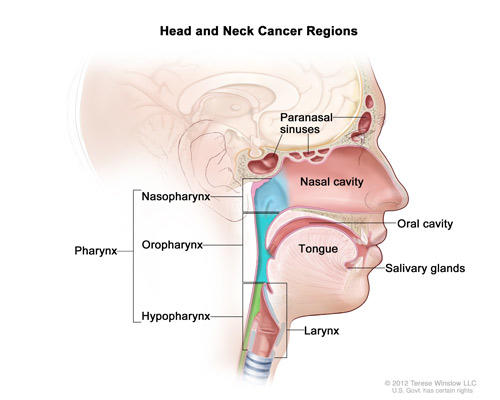}}
\caption{Head and neck cancer regions.}
\centering
\small Credit: © Terese Winslow
\label{fig1}
\end{figure}

ML models, such as random forests, gradient boosting, and deep learning, can process structured data and unstructured data to deliver personalized predictions. These models also help identify key factors influencing mortality risk, supporting informed clinical decisions \cite{b4, b5}.

In 2023, Dhariwal et al. presented a study on mortality predictions in HNSCC patients. This paper presents a novel approach to predicting mortality in head and neck cancers, focusing on the correlation between lifestyle factors like smoking, tobacco use, and key cancer attributes such as tumor-node-metastasis (TNM) staging and human papillomavirus (HPV) positivity. Using eight machine learning and four deep learning models, the study achieved a maximum accuracy of 98.8\% with XGBoost. The duration of follow-up emerged as the most influential factor in mortality prediction, contributing 40.8\% to the model's performance. The results, with a maximum area under the receiver operating characteristic (ROC) curve of 0.99. These studies further demonstrate the potential that ML models hold in predicting mortality in cancers, gradient boosting models predicting with high accuracies. \cite{b6, b7}. 

Kazmierski et al. \cite{b8} investigated prognostic modeling for head and neck cancer using deep learning and radiomics on data from 2552 patients. The multitask learning model provided the highest accuracy for predicting lifetime survival. Of the three models compared, the clinical data model performed best with an AUC of 0.74, emphasizing the importance of clinical features. The authors emphasized that while machine learning models show promise. They also suggested that ensemble learning techniques could enhance prediction accuracy.

Relapse-free survival (RFS) prediction estimates the likelihood that a cancer patient will remain free from recurrence after treatment. It is crucial for assessing treatment success and guiding long-term care decisions. Predictive models can help predict the relapse of cancer in patients based on their clinical and treatment data. this allows both doctors and patients to plan early for a possible relapse and prevent any fatalities.

HNSCC is a common cancer with a 5-year survival rate of about 50\%. While radiotherapy is a key treatment, predicting outcomes is vital for selecting effective therapies. Attributes like TNM, HPV status, and radiology expression show promise in predicting treatment response. However, reliable predictions for improving treatment are still needed \cite{b9}.

A predictive machine learning algorithm was developed to predict locoregional relapse at 18 months for oropharyngeal cancers with negative HPV status. The model, using clinical and Pyradiomics features from CT scans, was trained on the HN1 cohort (79 patients) and validated on the ART ORL cohort (45 patients). The XGBoost model achieved a precision of 0.92, recall of 0.42, AUC of 0.68, and accuracy of 0.64. Key features included voxel volume, grey level size zone matrix, and patient demographics (sex, age). This interpretable model shows potential for predicting relapse and guiding treatment decisions in clinical settings.

A predictive machine learning algorithm was proposed by Giraud et al. to predict locoregional relapse at 18 months for oropharyngeal cancers with negative HPV status. The XGBoost model, using clinical achieved a precision of 0.92, recall of 0.42, AUC of 0.68, and accuracy of 0.64. Key features included clinical features, medical insights, and patient demographics (sex, age). This interpretable model shows potential for predicting relapse and guiding treatment decisions in clinical settings \cite{b10}.

Freedom from distant metastasis (FDM) refers to the absence of cancer spread to distant organs after treatment, serving as a key indicator of prognosis. Predicting FDM helps identify high-risk patients for more intensive treatment or monitoring \cite{b11, 12}.

After careful study of literature, clinical features have proven to be highly effective in predicting mortality and chances of repalse on HNSCC patients. Further, it is also seen that most state-of-the-art models employ a highly trained XGBoost classifier due to being effective for prediction using clinical features because it efficiently handles complex, high-dimensional data, captures non-linear relationships, and offers high accuracy through feature importance ranking and regularization.

The aim for this study is to amalgamate clinical features and ML principles to present highly robust classifiers for precision oncology. The focused aims for the research are as follows:
\begin{enumerate}
    \item Identify and train a ML model, for HNSCC patients, to predict:
    \begin{itemize}
        \item Mortality;
        \item Cancer relapse free survival.
    \end{itemize}
    \item Identify the features that highly impact the predictions.
    \item Evaluate the performance of the classifiers and substantiate results with mertices. 
\end{enumerate}

\section{Methodology}

\subsection{Data Collection and Preprocessing}
The dataset for this research was sourced from the Cancer Imaging Archive by the National Cancer Institute, NIH \cite{b13}. The dataset consists of imaging, radiation therapy, and clinical data from head and neck squamous cell carcinoma (HNSCC) patients at MD Anderson Cancer Center, used in two research projects. The second project, "Radiomics Outcome Prediction in Oropharyngeal Cancer," focuses on integrating quantitative imaging biomarkers into current risk stratification tools. It uses clinical data and contrast-enhanced CT scans from 495 oropharyngeal cancer (OPC) patients treated between 2005 and 2012. The project aims to correlate radiomics features, either alone or in combination with clinical factors, with tumor outcomes. Radiomics analysis was conducted using institution-developed software on the Matlab platform.

\begin{figure*}[htbp]
\centerline{\includegraphics[width=120mm,scale=0.7]{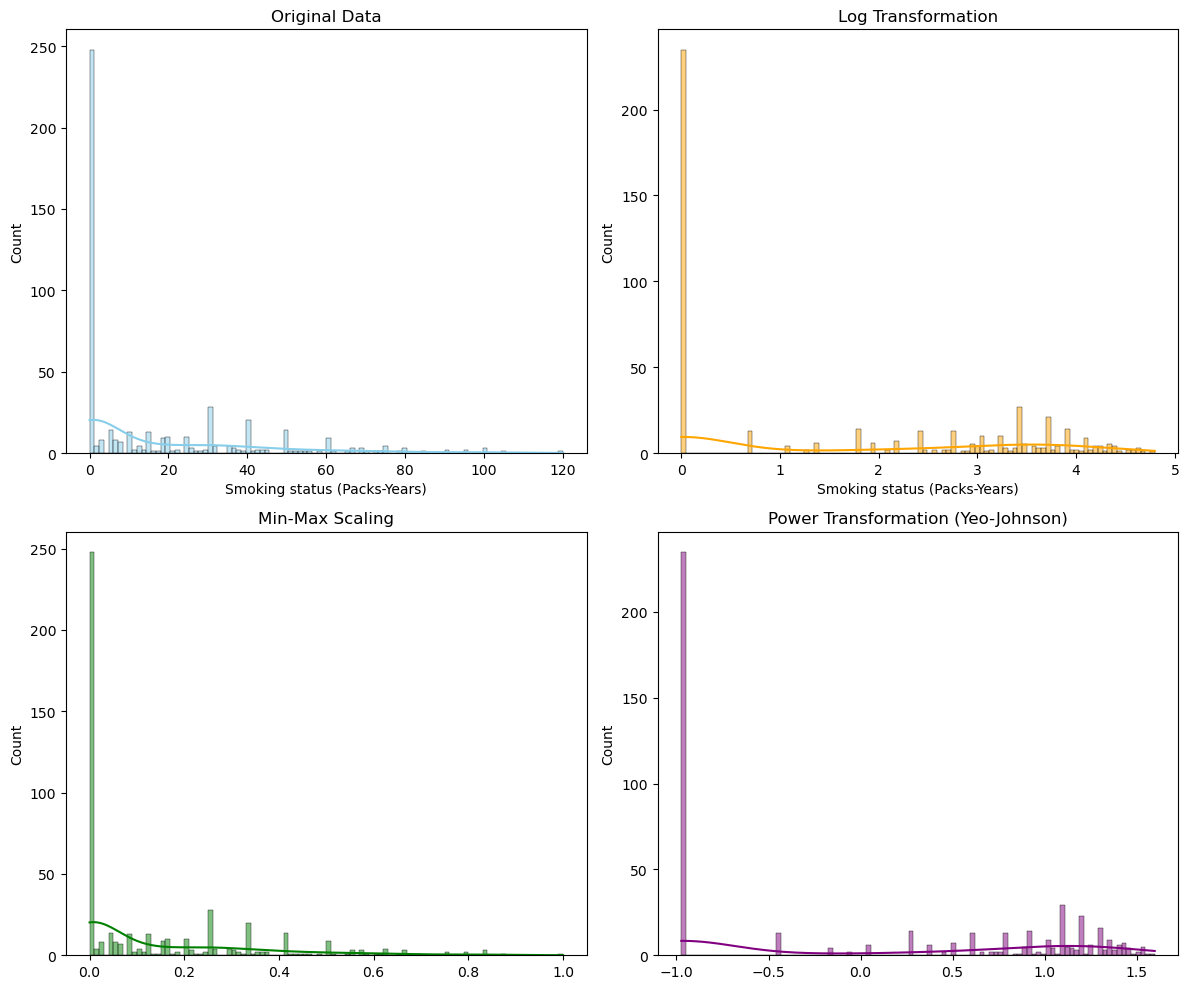}}
\caption{Comparison of Different Transformations Applied to the Smoking Status (Packs-Years) Feature}
\centering
\label{fig1}
\end{figure*}

The clinical dataset comprises of both numerical and categorical values. The data also contains a few missing values, which needed to be handled properly. One-hot encoding is used to convert categorical variables into numerical format by creating binary columns for each category. In this dataset, variables like "cancer stage," "gender," or "treatment type" are categorical. One-hot encoding transforms these variables into separate columns, where each category is represented by a 1 (present) or 0 (absent). For instance, the "cancer stage" with values "Stage I", "Stage II", "Stage III" and "Stage IV" would be transformed into 4 binary columns. This enables the model, such as XGBoost, to process categorical data without assuming any inherent order or numerical value, ensuring better model accuracy and interpretability.

To further prepare the dataset for modeling, normalization and standardization were applied to the numerical features based on their distributions. Predictors such as radiation treatment duration, total prescribed radiation treatment dose, and radiation treatment dose per fraction were standardized using the Robust Scaler due to the presence of outliers, ensuring that the influence of extreme values on these features was minimized. Additionally, the smoking status (packs-years) feature exhibited a highly skewed distribution that required transformation for improved model performance. Among the methods evaluated, the Yeo-Johnson transformation proved to be the most effective, outperforming both log transformation and min-max scaling.

The effectiveness of the Yeo-Johnson transformation is evident for two key reasons. First, it significantly reduced the skewness of the smoking status feature, as illustrated in the image, resulting in a distribution that is closer to symmetry. This is crucial because symmetric distributions tend to align better with the assumptions of many machine learning models, improving their predictive accuracy. Second, the Yeo-Johnson transformation natively handles zero and negative values without requiring any adjustments, unlike the log transformation, which necessitates adding a small constant to avoid undefined values for zeros. Given that the smoking status feature contains a significant proportion of zero values, this property of the Yeo-Johnson transformation makes it a more robust and seamless option.

\begin{figure*}[htbp]
\centerline{\includegraphics[width=110mm,scale=0.5]{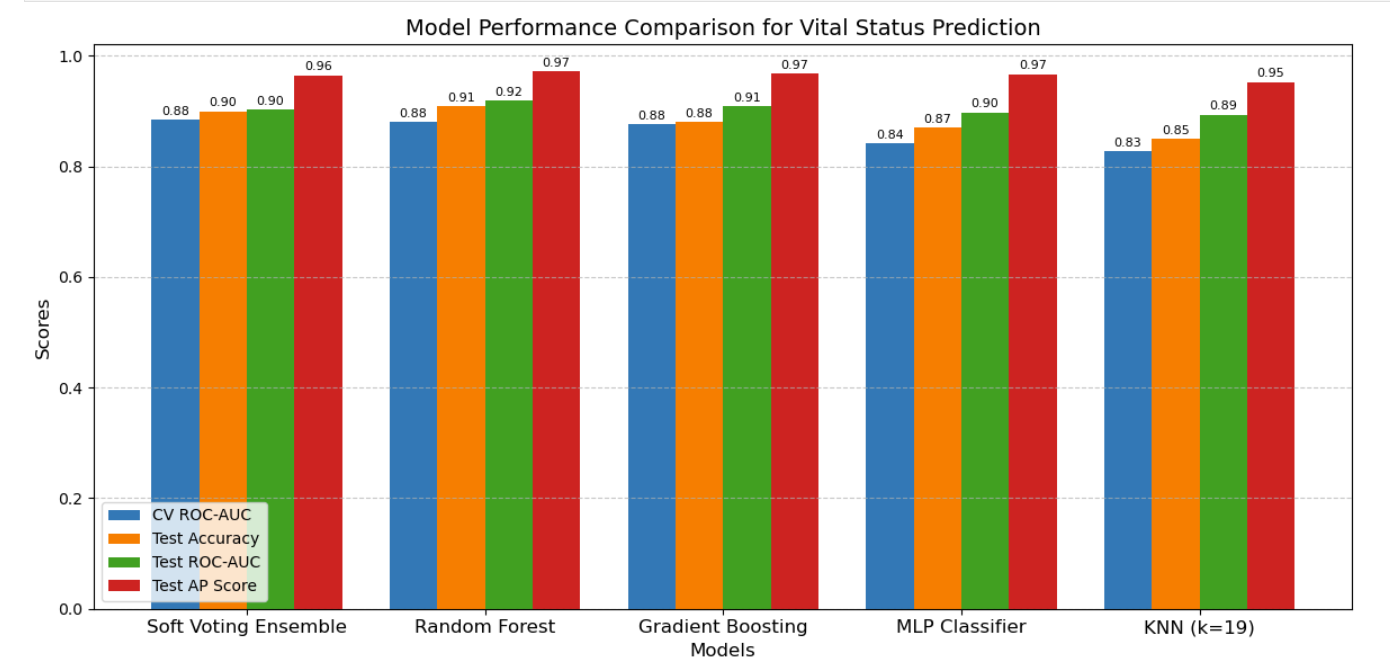}}
\caption{Classifier Comparison for mortality prediction.}
\centering
\label{cc}
\end{figure*}

Min-Max Scaling was applied to features such as 'T-category', 'N-category', 'Local control', and various duration-related variables to normalize them to a fixed range [0, 1], preserving their relative importance while ensuring compatibility with algorithms sensitive to scale, such as neural networks. To prevent data leakage, the dataset was split into training and test sets before applying any scalers or transformations. By incorporating these preprocessing steps, the pipeline ensures robust, efficient, and reliable model performance when predicting the target variable.

K-Folds cross-validation was employed instead of a traditional 80-20 data split. This technique involves dividing the dataset into 'k' subsets (or folds), and training the model on 'k-1' folds while testing it on the remaining fold \cite{b14}. It provides a more reliable estimate of model performance, especially when working with smaller datasets, and is beneficial for assessing the generalizability of the model and reducing overfitting.

\subsection{Model Configuration}

\subsubsection{Mortality Prediction}

An ensemble learning strategy that combines multiple classification algorithms was used. The dataset was preprocessed and split into training and testing sets using an 80-20 ratio, with stratification to maintain class distribution. Feature scaling was performed using standardization to ensure all variables contributed proportionally to the model predictions.

The ensemble model architecture consisted of two primary classifiers: Random Forest and Support Vector Machine (SVM). The Random Forest classifier was configured with 200 estimators, balanced class weights, and a maximum sampling rate of 0.7, incorporating a minimum sample split of 5 and minimum samples leaf of 4 to prevent overfitting. The SVM component utilized a linear kernel with C=0.1 and scaled gamma, optimized for probabilistic predictions to enable soft voting integration.

Multiple alternative classification algorithms were evaluated during the development phase, including Logistic Regression, Multi-Layer Perceptron (MLP), and K-Nearest Neighbors (KNN). However, preliminary cross-validation results indicated superior performance from the Random Forest and SVM combination, leading to their selection for the final ensemble. The model development process included extensive hyperparameter tuning using grid search cross-validation to optimize each algorithm's parameters.

The ensemble implemented a soft voting strategy, combining probability predictions from both constituent models rather than hard class assignments. This approach allows for more nuanced decision-making by considering the confidence levels of individual classifiers. Model evaluation was conducted using 5-fold stratified cross-validation to ensure robust performance assessment, with metrics including ROC-AUC, precision, recall, and F1-score to provide a comprehensive evaluation of predictive capabilities.

\subsubsection{Relapse Prediction}

In this study, three core machine learning supervised learning algorithms; XGBoost, Random Forest and Support Vector Machines; were configured to predict "Relapse-free survival" based on clinical features. The target is a binary column with classes "Relapse" and "No relapse" which have been converted to 1 and 0 respectively for model calculations. The target variable, "Relapse-free survival," was separated from the feature set, and the data was split into training and testing sets using an 80-20 ratio, followed by K-folds cross validation. The data was then normalized and scaled appropriately according to the pre-processing discussed in previous section. Each model was then hyper-parameter tuned and experimented with different configurations to find the best performance and recall balance.

After fitting the model to the training data, performance metrics were computed, including precision, recall, accuracy and Area Under the Curve (AUC) for the data and classes. The AUC scores were calculated using the model's predicted probabilities, which provide a more nuanced evaluation of the model's ability to discriminate between positive and negative outcomes.

Recall was prioritized in this study to minimize false negatives, which is critical in medical predictions where undiagnosed cases can lead to delayed treatment and adverse outcomes. Emphasizing recall ensures the model identifies all potential cases, aligning with the clinical objective of improving patient safety and outcomes.

Feature importance was extracted from the trained model, which provides insights into which features contribute most significantly to the prediction of relapse-free survival. The importance scores were used to rank the features in descending order, highlighting the most influential variables.

Additionally, model evaluation was conducted using Receiver Operating Characteristic (ROC) curves to visualize the model's classification performance. The learning curve was also plotted to assess how the model's performance evolves with increasing training data, providing an indication of the model’s capacity to generalize across different dataset sizes.

This configuration ensures robust evaluation and understanding of the model's predictive capabilities in the context of relapse-free survival, leveraging the power of XGBoost's, Random Forest's and Support Vector Machine's gradient-boosting approach to handle complex relationships within the clinical data.

The XGBoost model is used as a baseline for the Random Forest and Support Vector Machine classifier, based on the existing studies, to provide a comparative foundation for evaluating performance. Random Forest, an ensemble learning method based on multiple decision trees, offers a strong baseline due to its simplicity, robustness, and ability to handle complex datasets. Support Vector Machines (SVM) were employed in this study due to their effectiveness in handling high-dimensional data and their robust performance in binary classification tasks. SVM's ability to maximize the margin between classes ensures better generalization, making it particularly suitable for medical datasets with complex, non-linear relationships.

\section{Results and Discussions}

\subsection{Mortality Prediction}
The ensemble model demonstrated robust predictive performance, achieving a mean cross-validation ROC-AUC score of 0.884 (±0.062) and an overall test accuracy of 90\%. This performance was notably superior to individual model implementations, with the Random Forest achieving a mean cross-validation ROC-AUC of 0.880 (±0.074) and the Gradient Boosting classifier scoring 0.876 (±0.063). The ensemble's balanced performance across both classes, with precision scores of 0.84 and 0.91 for classes 0 and 1 respectively, indicates its effectiveness in handling potential class imbalance issues.

Analysis of feature importance derived from the Random Forest component revealed that "Locoregional control duration," "Days to last FU," and "Local control duration" were the most influential predictors, collectively accounting for approximately 47\% of the total feature importance. This finding provides valuable insights into the key clinical factors driving vital status predictions and aligns with clinical expertise in the field. The identification of these crucial features can help guide clinical decision-making and resource allocation in patient care.

\begin{table*}[ht]
\centering
\caption{Model Performance Comparison for Vital Status Prediction}
\begin{tabular}{|c|c|c|c|c|c|c|}
\hline
\textbf{Model} & \textbf{CV ROC-AUC} & \textbf{Test Accuracy} & \textbf{Test ROC-AUC} & \textbf{Test AP Score} & \textbf{Precision (0/1)} & \textbf{Recall (0/1)} \\ \hline
Soft Voting Ensemble & 0.884 ±0.062 & 0.90 & 0.903 & 0.965 & 0.84/0.91 & 0.70/0.96 \\ \hline
Random Forest & 0.880 ±0.074 & 0.91 & 0.919 & 0.972 & 0.82/0.94 & 0.78/0.95 \\ \hline
Gradient Boosting & 0.876 ±0.063 & 0.88 & 0.910 & 0.969 & 0.79/0.90 & 0.65/0.95 \\ \hline
MLP Classifier & 0.842 ±0.043& 0.87 & 0.898 & 0.967 & 0.73/0.91 & 0.70/0.92 \\ \hline
KNN (k=19) & 0.827 ±0.031 & 0.85 & 0.894 & 0.953 & 1.00/0.84 & 0.35/1.00 \\ \hline
\end{tabular}
\label{vital_status_model_eval}
\end{table*}

The confusion matrix analysis revealed strong performance in identifying both positive and negative cases, with only 10 misclassifications out of 99 test samples. The model showed particularly high specificity, correctly identifying 73 out of 76 positive cases, while maintaining reasonable sensitivity with 16 correct identifications out of 23 negative cases. This performance pattern suggests the model is well-suited for clinical applications where both false positives and false negatives carry significant implications.

\begin{figure*}[htbp]
\centerline{\includegraphics[width=100mm,scale=0.5]{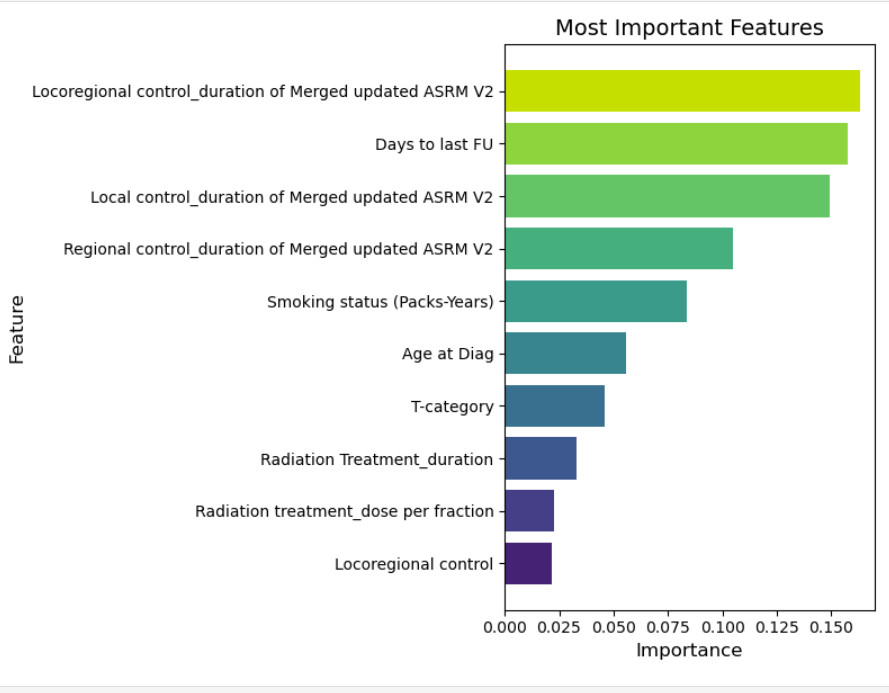}}
\caption{Most influential features for the prediction of mortality in patients.}
\centering
\label{mfi}
\end{figure*}

The superior performance of the ensemble compared to alternative algorithms (including Logistic Regression, MLP, and KNN) validates the chosen architecture and highlights the benefits of combining complementary classification approaches. The soft voting strategy effectively leveraged the strengths of both Random Forest's ability to capture non-linear relationships and SVM's effective boundary optimization. Moreover, the stability of cross-validation scores across folds (standard deviation of ±0.062 in ROC-AUC) indicates robust generalization capabilities, suggesting the model would maintain reliable performance when deployed in clinical settings. A visual comparison of the models and their performances is shown in figure \ref{cc}. Further, figure \ref{mfi} visualized and summarizes the most influential features for predicting the mortality in HNSCC patients.

\subsection{Relapse Prediction}

Table~\ref{RFS_model_eval} presents the performance metrics of various machine learning models in predicting relapse outcomes. The evaluation criteria include precision, recall, accuracy, and ROC-AUC scores, with results reported for both class 0 (no relapse) and class 1 (relapse). Among the models tested, Support Vector Machine (SVM) 2 exhibited the most balanced performance across metrics. It achieved a precision of 0.94 and 0.97 for classes 0 and 1, respectively, coupled with a recall of 0.88 and 0.99, highlighting its reliability in minimizing false negatives—a critical requirement in medical diagnosis. The accuracy of 96\% and a competitive ROC-AUC of 0.9288 further demonstrate SVM 2's capability to differentiate between relapse and non-relapse cases effectively. This superior recall and ROC-AUC performance make SVM 2 particularly suitable for medical applications, where the cost of misdiagnosis is high and ensuring the correct identification of at-risk patients is paramount.

While Random Forest also delivered strong results, with its best configuration reaching an accuracy of 96\% and ROC-AUC of 0.9360, its slightly lower recall for class 0 (0.82) indicates a potential limitation in identifying non-relapse cases compared to SVM 2. XGBoost models demonstrated high accuracy (up to 94\%) and competitive ROC-AUC values (up to 0.9371), but their lower recall for class 0 (0.78) in some cases suggests reduced sensitivity in identifying non-relapse cases, which is less favorable in medical applications.

\begin{figure*}[htbp]
\centerline{\includegraphics[width=110mm,scale=0.5]{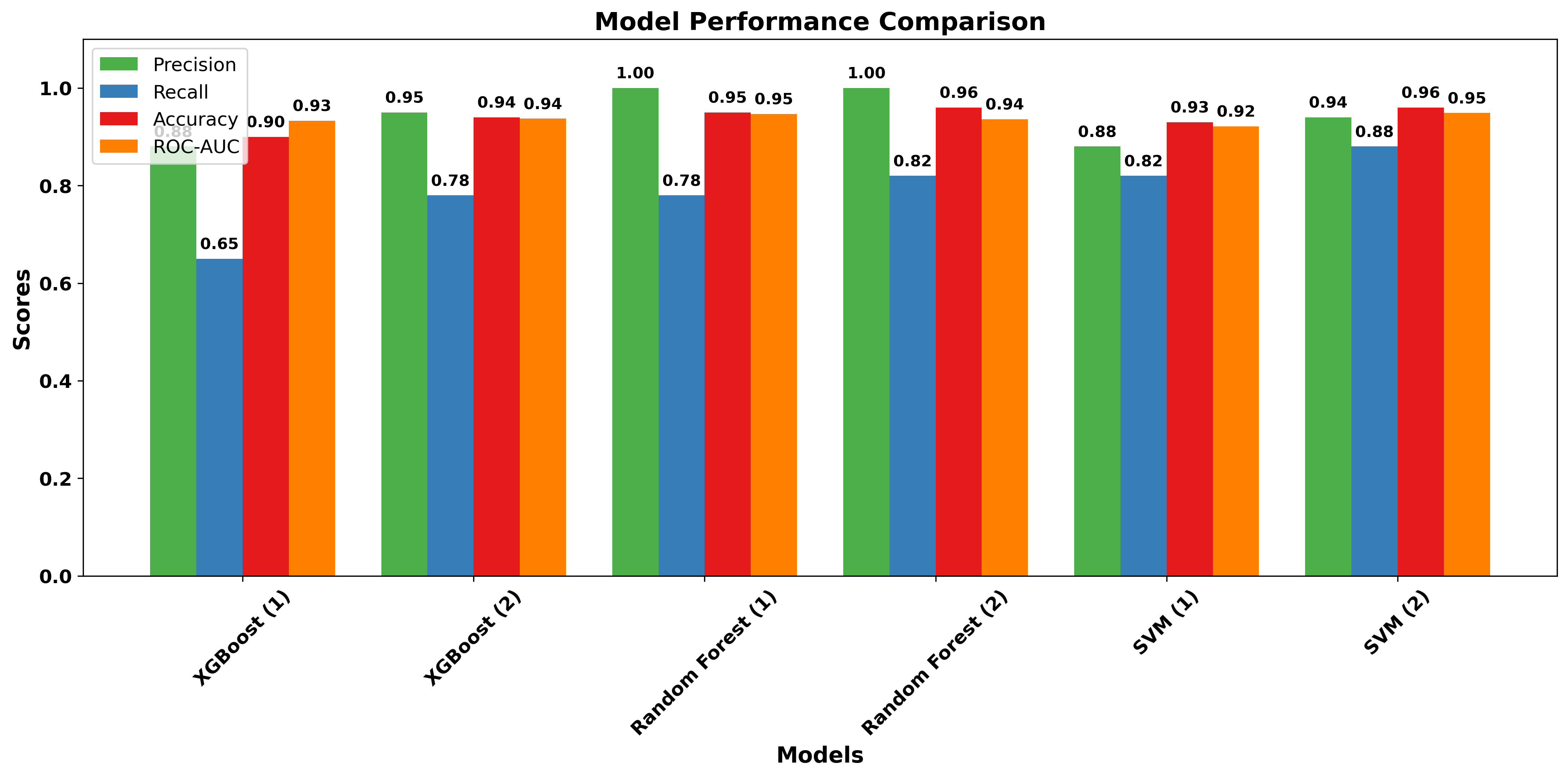}}
\centering
\caption{Visualization of classifiers; performance for predicting relapse free survival.}
\label{model_comp_rfs}
\end{figure*}

SVM 2 consistently achieved high recall and precision for both classes, making it the most reliable model for relapse prediction. The ability to identify at-risk patients (class 1) with near-perfect recall (0.99) while maintaining high precision underscores its suitability for medical diagnosis, where the cost of false negatives is particularly significant. The results have been visualized in figure \ref{model_comp_rfs}.

\begin{table*}[t]
\centering
\caption{Model Performance Comparison for Relapse Prediction}
\begin{tabular}{|c|c|c|c|c|}
\hline
\textbf{Model}                 & \textbf{Precision (Class 0/Class 1)} & \textbf{Recall (Class 0/Class 1)} & \textbf{Accuracy} & \textbf{ROC-AUC} \\ \hline
XGBoost                        & 0.88/0.90                           & 0.65/0.97                        & 0.90              & 0.9331           \\ \hline
XGBoost                        & 0.95/0.94                           & 0.78/0.99                        & 0.94              & 0.9371           \\ \hline
Random Forest                  & 1.00/0.94                           & 0.78/1.00                        & 0.95              & 0.9468           \\ \hline
Random Forest                  & 1.00/0.95                           & 0.82/1.00                        & 0.96              & 0.9360           \\ \hline
Support Vector Machine         & 0.88/0.95                           & 0.82/0.96                        & 0.93              & 0.9216           \\ \hline
Support Vector Machine         & 0.94/0.97                           & 0.88/0.99                        & 0.96              & 0.9288           \\ \hline
\end{tabular}
\label{RFS_model_eval}
\end{table*}

\begin{figure*}[htbp]
\centerline{\includegraphics[width=105mm,scale=0.5]{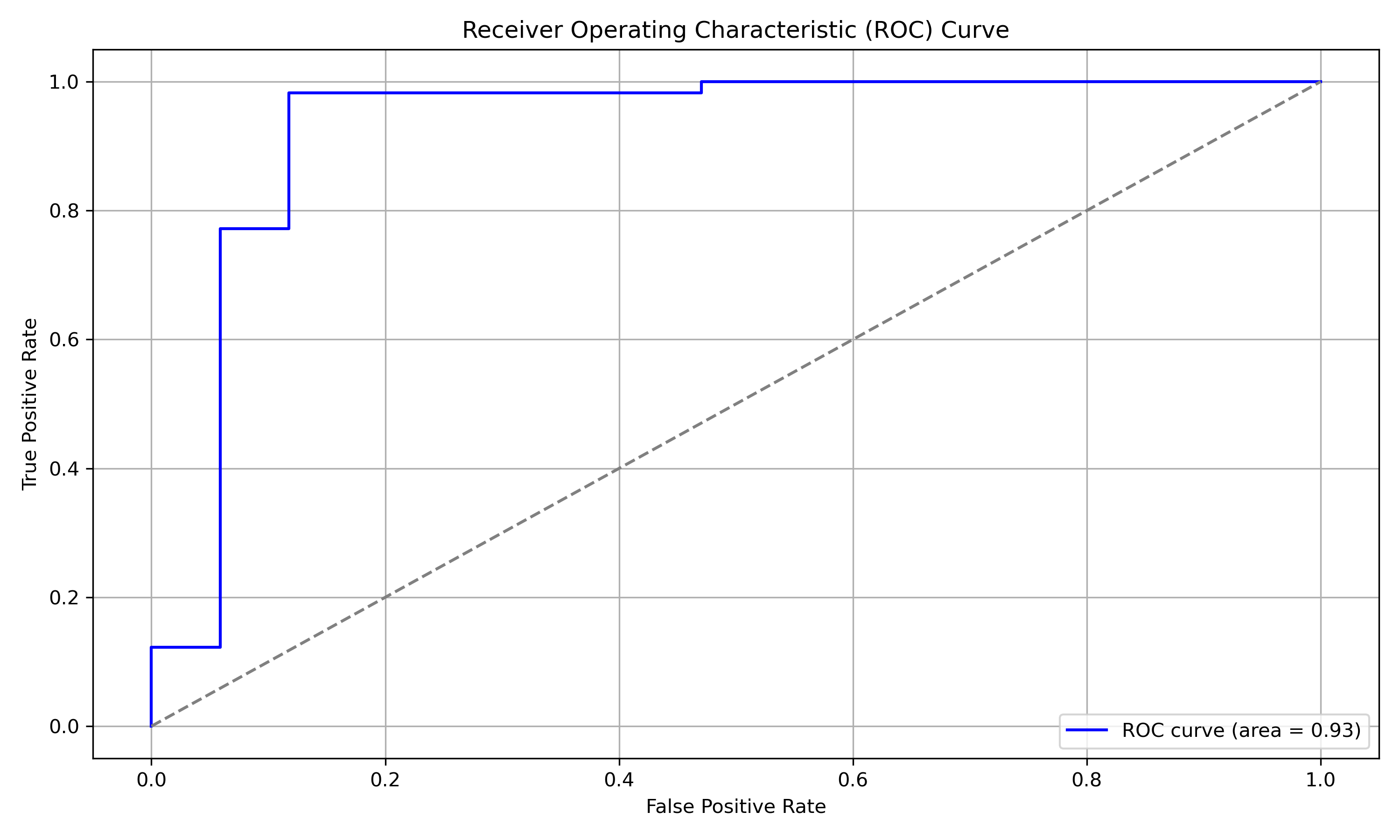}}
\centering
\caption{ROC Curve for Support Vector Machine for predicting relapse free survival.}
\label{roc_rfs_xg}
\end{figure*}

The superior performance of Support Vector Machine (SVM) 2 in predicting relapse can be attributed to several inherent strengths of the SVM algorithm, particularly its ability to achieve a balanced trade-off between precision and recall. SVM is a robust classifier that excels in high-dimensional spaces, making it particularly suited for clinical data, where the relationships between input features can be complex and non-linear. The ability of SVM 2 to maintain high precision (0.94/0.97) while achieving near-perfect recall (0.88/0.99) across both classes suggests that the model is adept at minimizing both false positives and false negatives, which is of paramount importance in medical contexts. In healthcare, particularly in relapse prediction, the cost of false negatives (failing to detect a relapse) is significantly higher than that of false positives (misclassifying a non-relapse as a relapse), making high recall for class 1 (relapse) particularly critical.

\begin{figure*}[htbp]
\centerline{\includegraphics[width=110mm,scale=0.5]{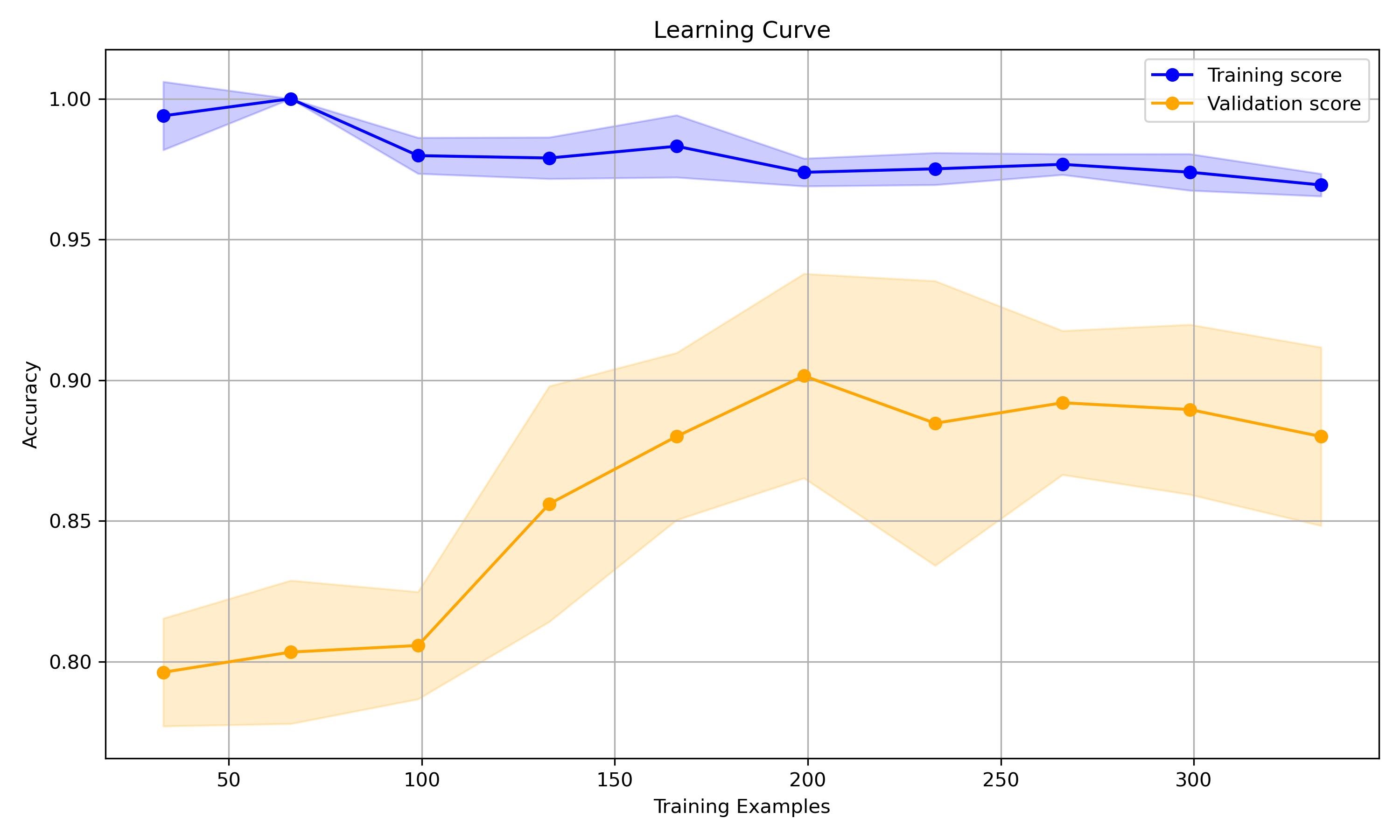}}
\centering
\caption{Learning Curve for Support Vector Machine for predicting relapse free survival.}
\label{roc_rfs_rf}
\end{figure*}

Moreover, SVM’s kernel trick, which enables it to efficiently map input features into higher-dimensional spaces, might allow it to capture intricate patterns in the data that other models, such as Random Forest and XGBoost, may miss. This could explain the consistently high recall for class 1 and the overall high accuracy of 96\%. The model’s ability to achieve high ROC-AUC (0.9288) further suggests its strong discriminatory power between the classes, which is essential for distinguishing between relapse and non-relapse cases.

On the other hand, while Random Forest and XGBoost demonstrated competitive performance with high accuracy and ROC-AUC scores, their slightly lower recall for class 0 (non-relapse) indicates that these models might be less sensitive in identifying non-relapse cases. This limitation could arise from their inherent tendency to favor overall accuracy or from the nature of the data, where class 0 may be underrepresented or more difficult to differentiate. Random Forest, which relies on ensemble learning from multiple decision trees, and XGBoost, which optimizes decision trees using gradient boosting, both benefit from the aggregation of multiple weak learners. However, they might struggle with balancing sensitivity (recall) and specificity in certain contexts, particularly when class imbalance is present or when fine-tuning hyperparameters for balanced classification is challenging.

\begin{figure*}[htbp]
\centerline{\includegraphics[width=100mm,scale=0.5]{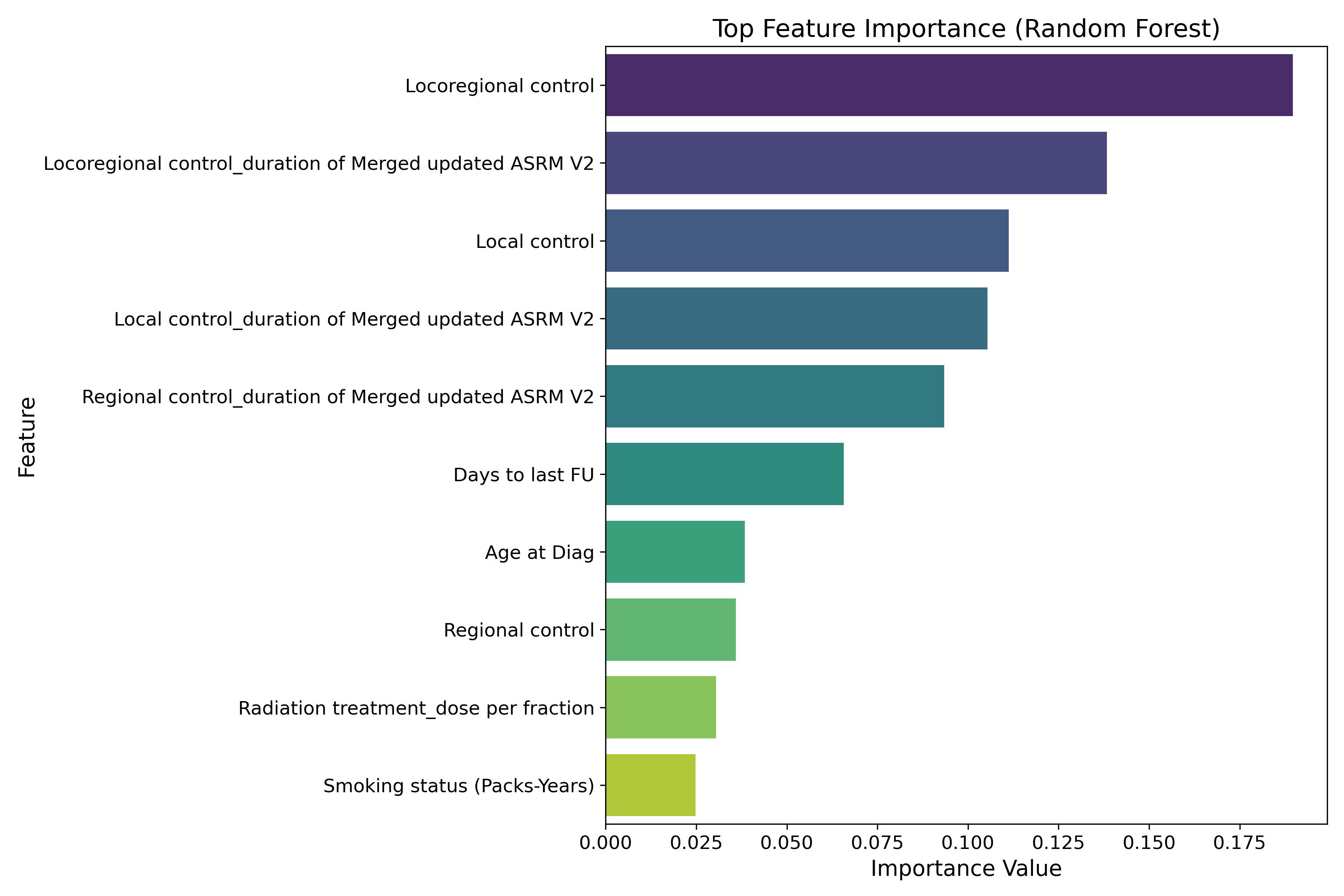}}
\caption{Feature Importance from Random Forest Classifier (2) for predicting relapse free survival.}
\centering
\label{rfs_feat_imp}
\end{figure*}

The Random Forest model reveals that locoregional control is the most influential feature, indicating its strong association with relapse-free survival. Other notable features include control factors, days to last treatment, and smoking history, which contribute to understanding patient prognosis and survival outcomes. The top 10 most influencial feature for the XGBoost classifier are visualized with their impact score in figure \ref{rfs_feat_imp}.

In the real-world context, these results highlight the importance of monitoring locoregional control and distant metastasis in clinical settings, as these factors are strongly linked to patient survival and relapse-free periods. Furthermore, factors such as HPV status and smoking history emphasize the need for personalized treatment strategies, focusing on these key risk factors to optimize patient care.

\section{Conclusion}

The findings of this study underscore the transformative potential of machine learning (ML) in predicting clinical outcomes for patients with head and neck squamous cell carcinoma (HNSCC). By leveraging advanced ML techniques on clinical data, the study achieved high predictive accuracies across key prognostic tasks, including mortality and relapse-free survival. The ensemble soft-voting strategy demonstrated the advantage of combining diverse models for robust mortality predictions, while SVM outperformed other models in capturing the nuances of relapse-free survival.

Through feature importance analysis, this research identified critical clinical factors, such as locoregional control, regional control, tumor stage, treatment methods and smoking. These insights provide actionable value for patient stratification, aiding clinicians in optimizing treatment plans and improving care outcomes. The study establishes a framework for integrating ML-driven insights into precision oncology, facilitating early interventions and more effective, personalized care pathways for HNSCC patients.

Future research should aim to build on these promising results by incorporating multimodal data, such as genomic and radiomic features, to enhance predictive robustness and clinical applicability. Expanding the data set to include larger multiinstitutional cohorts will further validate the findings and support broader adoption in clinical settings. By bridging the gap between computational advances and medical practice, this work represents a critical step towards enhancing prognostic precision and improving the quality of life for cancer patients.


\begin{thebibliography}{00}
\bibitem{b1} Siegel, Rebecca L., Angela N. Giaquinto, and Ahmedin Jemal. "Cancer statistics, 2024." CA: a cancer journal for clinicians 74.1 (2024): 12-49.
\bibitem{b2} Chow, Laura QM. "Head and neck cancer." New England Journal of Medicine 382.1 (2020): 60-72.
\bibitem{b3} Mody, Mayur D., et al. "Head and neck cancer." The Lancet 398.10318 (2021): 2289-2299.
\bibitem{b4} Gormley, Mark, et al. "Reviewing the epidemiology of head and neck cancer: definitions, trends and risk factors." British Dental Journal 233.9 (2022): 780-786.
\bibitem{b5} Patterson, Rolvix H., et al. "Global burden of head and neck cancer: economic consequences, health, and the role of surgery." Otolaryngology–Head and Neck Surgery 162.3 (2020): 296-303.
\bibitem{b6} Dhariwal, Naman, Rithvik Hariprasad, and L. Mohana Sundari. "An artificial intelligence based approach toward predicting mortality in head and neck cancer patients with relation to smoking and clinical data." IEEE Access 11 (2023): 126927-126937.
\bibitem{b7} Dhariwal, Naman. "Brain Metastasis Origin and Patient Mortality Predictions Using MRI with Clinical and Imaging Feature Information by Deep Learning Architectures." 2024 3rd International Conference for Innovation in Technology (INOCON). IEEE, 2024.
\bibitem{b8} Kazmierski, Michal, et al. "Multi-institutional prognostic modeling in head and neck cancer: evaluating impact and generalizability of deep learning and radiomics." Cancer Research Communications 3.6 (2023): 1140-1151.
\bibitem{b9} Begg, Adrian C. "Predicting recurrence after radiotherapy in head and neck cancer." Seminars in radiation oncology. Vol. 22. No. 2. WB Saunders, 2012.
\bibitem{b10} Giraud, Paul, et al. "Interpretable machine learning model for locoregional relapse prediction in oropharyngeal cancers." Cancers 13.1 (2020): 57.
\bibitem{b11} Duprez, Fréderic, et al. "Distant metastases in head and neck cancer." Head \& neck 39.9 (2017): 1733-1743.
\bibitem{b12} Ferlito, Alfio, et al. "Incidence and sites of distant metastases from head and neck cancer." ORL 63.4 (2001): 202-207.
\bibitem{b13} Grossberg A, Elhalawani H, Mohamed A, Mulder S, Williams B, White AL, Zafereo J, Wong AJ, Berends JE, AboHashem S, Aymard JM, Kanwar A, Perni S, Rock CD, Chamchod S, Kantor M, Browne T, Hutcheson K, Gunn GB, Frank SJ, Rosenthal DI, Garden AS, Fuller CD, M.D. Anderson Cancer Center Head and Neck Quantitative Imaging Working Group. (2020) HNSCC Version 4 [Dataset]. The Cancer Imaging Archive. DOI: https://doi.org/10.7937/k9/tcia.2020.a8sh-7363
\bibitem{b14} Anguita, Davide, et al. "The'K'in K-fold Cross Validation." ESANN. Vol. 102. 2012.
\bibitem{b15} Chen, Tianqi, and Carlos Guestrin. "Xgboost: A scalable tree boosting system." Proceedings of the 22nd acm sigkdd international conference on knowledge discovery and data mining. 2016.

\end{thebibliography}
\end{document}